\documentclass[a4paper,fleqn,usenatbib]{mnras}

\usepackage[T1]{fontenc}
\usepackage{ae,aecompl}
\usepackage[utf8]{inputenc}

\usepackage{graphicx}	
\usepackage{amsmath}	
\usepackage{amssymb}	
\usepackage[english]{babel}

\usepackage{newtxtext,newtxmath}

\usepackage[dvipsnames]{xcolor}
\newcommand{\myr}{$\rm mas\,yr^{-1}$}
\newcommand{\kms}{$\rm km\,s^{-1}$}
\newcommand{\kmskpc}{$\rm km\,s^{-1}\,kpc^{-1}$}

\definecolor{myviolet}{RGB}{192, 130, 255}

\title[Parameters of the spiral arms from kinematic tracers]
{Determining the parameters of the spiral arms of the Galaxy from kinematic tracers based on {\it Gaia} DR3 data}

\author[Denyshchenko et al.]{
S. I. Denyshchenko,$^{1}$\thanks{E-mail:sofiia.denyshchenko@gmail.com (SID)}
P. N. Fedorov,$^{1}$\thanks{E-mail:pnfedorov@gmail.com (PNF)}
V. S. Akhmetov,$^{1,2}$\thanks{E-mail:akhmetovvs@gmail.com (VSA)}  
A. B. Velichko$^{1,3}$  
and A. M. Dmytrenko$^{1}$
\\
$^{1}$Institute of astronomy of V. N. Karazin Kharkiv national university, Svobody sq. 4, 61022, Kharkiv, Ukraine\\
$^{2}$INAF-Osservatorio Astrofisico di Torino, Via Osservatorio 20, Pino Torinese, Turin, I-10025, Italy\\
$^{3}$Department of Astronomy, University of Geneva, Chemin Pegasi 51, 1290 Versoix, Switzerland
}

\date{Accepted XXX. Received YYY; in original form ZZZ}

\pubyear{2023}

\begin{document}
\label{firstpage}
\pagerange{\pageref{firstpage}--\pageref{lastpage}}
\maketitle

\begin{abstract}
We present the results of determining the parameters of the spiral arms of the Galaxy using the stars {\it Gaia} DR3, whose absolute magnitude is $M_G<$ 4, and which allow tracing spiral arms at large distances from the Sun. As tracers of spiral arms, we use the centroids of stellar spherical regions with a radius of 0.5 kpc, in which the deformation velocities along the coordinate axis $R$ are insignificant. These kinematic tracers cover the Galactic plane within the Galactocentric coordinate ranges $140^\circ < \theta < 220^\circ$ and 4 kpc $ < R < $ 14 kpc. The numerical values of the pitch angles of the spirals and their Galactocentric distances to the point of intersection of the spiral with the direction of the Galactic center - the Sun are in good agreement with the results of other authors. By extrapolating beyond the data we have, we present a schematic four-arm global pattern, consisting of the Scutum–Centaurus, Sagitarius-Carina, Perseus, Norma-Outer arms, as well as the local arm Orion. The uncertainties of the determined spiral parameters confirm that the structures identified are not false, but are reliable from the statistical point of view.
\end{abstract}


\section{Introduction}

The nature of the formation of the spiral structure of the galaxy is still not determined. There are many assumptions about the causes of arms and spiral patterns in the Galaxy \citep[][]{Toomre1981, Sellwood2002, Masset1997, Merrifield2006, Patsis2006, Quillen2011}

According to the classical wave theory, proposed by \citet{Lin1964}, spiral arms are waves of increased density rotating around the center of the Galaxy as a solid body at a constant angular velocity (i.e., pattern velocity), despite the differential rotation of stars and interstellar medium, that results in long lifetime of spiral arms \citep[see also][]{Shu2016}. To date, many works have been devoted to the study of the spiral structure, using various methods and objects \citep[][]{Lin1964, Lin1969, Georgelin1976, Taylor1993, Russeil2003, Paladini2004, Popova2005, Dias2005, Levine2006, Moitinho2006, Vazquez2008, Gerhard2011, Efremov2011, Bobylev2014, Hou2014, Hou2015, Dambis2015, Reid2019, Xu2018, Xu2021, Poggio2021, Hao2021}.
Most of these methods are based on the analysis of trigonometric parallaxes and radial velocities of young objects such as OB stars, open clusters, hydrogen clouds, HII regions, giant molecular clouds, methanol masers, etc. However, the spiral structures of the Galaxy contain not only young objects, but also most other types of objects, including red giants and subgiants. For example, \citet{Junqueira2015} used a sample of giants of different ages for a new method for determining the velocity of the Milky Way spiral pattern based on the interaction between spiral arms and disk stars.
The main problems in the studies given above are the difficulty in determining trigonometric parallaxes, their significant errors, as well as the limited number of samples of the objects under study.

The third release of the {\it Gaia} space mission, which contains high-precision proper motions, parallaxes, radial velocities and other astrometric and astrophysical parameters for stars of various types, makes it possible to analyze in great details the structural features of an extended region of our Galaxy. A large sample of objects can be used not only to analyze the spatial structure of the Galaxy from positions of the objects, but also to perform a detailed kinematic analysis of Galactic various regions, that makes it possible to obtain additional information about the structural features of these regions. A frequently used physical model for studying stellar kinematics is the Ogorodnikov-Milne (O--M) model \citep[][]{Ogorodnikov1965, Clube1972, DuMont1977, Miyamoto1993, Miyamoto1998}. This model makes it possible to study the velocity field in a deformable stellar system. \citet{Fedorov2021, Fedorov2023} proposed a method for obtaining kinematic parameters within the framework of the O--M model for stellar systems whose centroids are located in the Galactic plane. In these works, the region of determination of the kinematic parameters of stellar systems has been extended to heliocentric distances up to 10 kpc.

One of the main goals of the spiral structure theory, which is based on observational data, is to explain the stability of spiral patterns. Why do these patterns exist throughout the many revolutions of the Galaxy, while the differential galactic rotation tends to destroy these patterns? Obviously, if the spiral pattern existed for a short period of time, we would not observe them in a large number of disk galaxies, since differential rotation deforms any structure in the disk, causing it to be destroyed during one or two rotations of the disk. One of the possible explanations for the stability of spiral patterns is the absence of contraction-expansion deformations that destroy their shape in those regions of the Galaxy where spiral arms are located.

From a kinematic point of view, this means that in the field of stellar velocities described by the O--M model, the diagonal components of the deformation velocity tensor should be close to zero. In this case, the deformation velocities of the stellar system along the coordinate axes (relative contraction-expansion velocities) will be small on average or will be completely absent, which probably affects the stability of the stellar structure configuration. The results of determining the parameters of spiral arms given in this work are based on the assumptions mentioned above and are verified by comparing them with the results obtained in various works by other authors.

\section{Kinematic parameters and data used}

In our previous papers \citep[][]{Fedorov2021, Fedorov2023, Dmytrenko2023}, the kinematic parameters of the O--M model for stellar systems contained in spherical regions of a 1 kpc radius, whose centroids are located in the Galactic plane, have been determined in local coordinate systems. The orientation of the local coordinate system with the origin at an arbitrary point of the Galactic plane corresponds to the following conditions: the $x'$ axis is directed to the center of the Galaxy, the $y'$ axis is in the direction of the Galaxy rotation, and the $z'$ axis is perpendicular to the Galactic plane. As shown earlier \citep[][]{Dmytrenko2023}, in the local coordinate system, the components $M^+_{13}$ and $M^+_{23}$ of the velocity deformation tensors for these stellar systems are close to zero over the entire range of changes in the Galactocentric cylindrical coordinates $R$ and $\theta$, which indicates the motion of stars, mainly in the Galactic plane. Therefore, in this work, we consider the tensor $M^+$ as flat, i.e., having only four components $M^+_{11}, M^+_{12}, M^+_{21}, M^+_{22}$, of which only three are independent ($M^+_{12} = M^+_{21}$).

\begin{figure}
\centering
\resizebox{\hsize}{!}
   {\includegraphics{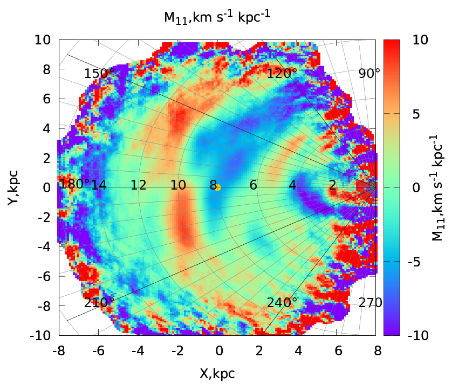}}
  \caption{The O--M model parameter $M^+_{11}$ taken from \protect\citet{Fedorov2023} as a function of the Galactic coordinates.}
\label{fig:M11}
\end{figure}

\begin{figure}
\centering
\resizebox{\hsize}{!}
   {\includegraphics{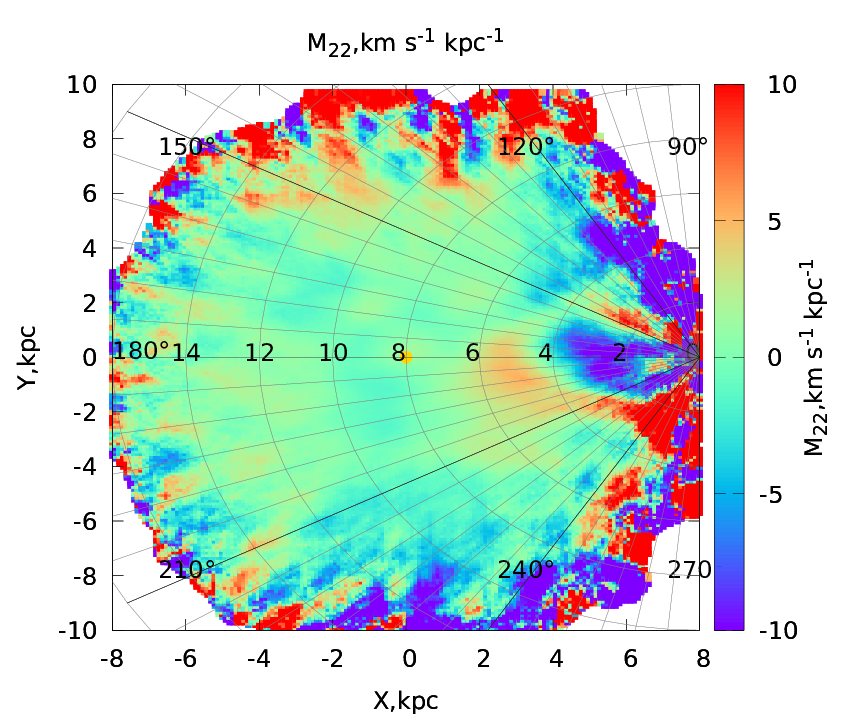}}
  \caption{The O--M model parameter $M^+_{22}$ taken from \protect\citet{Fedorov2023} as a function of the Galactic coordinates.}
\label{fig:M22}
\end{figure}

Figures \ref{fig:M11}, \ref{fig:M22} show the diagonal components of the deformation velocity tensor $M^+_{11}$ and $M^+_{22} $, which have been obtained in our previous work \citep[][]{Fedorov2023}. As can be seen from Fig. \ref{fig:M22}, in the range of angles $\theta$ from 210 to 150 degrees and distances $R$ from 4 to 12 kpc, the magnitude of change in velocity $V_\theta$ along the unit vector \textbf{\textit j} (parameter $M_{22}$), that characterizes contraction - expansion along the $y'$ axis of the local coordinate system is approximately $\pm$2.5\kmskpc. Compared to the average velocity $V_\theta$ (about 240 \kms), this is approximately from 0.5 to 12\% depending on $R$. At the same time, in the same region of the Galaxy, the $M^+_{11}$ parameter, which contraction - expansion along the $x'$ axis of the system (or along the unit vector \textbf{\textit i}), reaches values of about $\pm$10 \kmskpc\, inside some ring-shaped structures. Compared to the radial velocity $V_R$ (about $\pm$10 \kmskpc), its change $\partial V_R/\partial R$ = $M^+_{11}$ is almost 100\% depending on $R$ and is decisive in the formation of velocity field deformations. In this regard, for the region of the Galaxy under study, we neglect the influence of the velocity deformation along the $y'$ axis of the local coordinate system on the stability of the stellar structure configuration.

Also, Fig. \ref{fig:M11} clearly shows that next to these ring-shaped structures in the Galactic plane there are regions adjacent, inside which the gradient $\partial V_R/\partial R$ = $M^+_{11}$ is close to zero. For example, this is a ring-shaped structure located at a Galactocentric distance of about 13 kpc. Similar structures, although less clear, can be seen at distances of about 9 kpc, 6 kpc, and 4 kpc. This means that the velocities of the deformation velocities along the $R$ direction inside these regions are about zero, as a result of which there is no expansion or contraction of stellar systems along the $R$ coordinate. Then in the case when its average value is very close to zero, there should be no deformations of the structures observed in these regions, and as a result, their shape should not change with time. The property of the invariability of the shape of structures containing objects with the kinematic parameter $\partial V_R/\partial R \approx 0$ allows us to assume that there is a relation between the stability of the spiral pattern in our Galaxy and the parameter $\partial V_R/\partial R \approx 0$.

\subsection{Sample}

To test the assumption mentioned above, we created a special sample of stars from the third release of the {\it Gaia} data -- {\it Gaia} DR3 \citep[][]{Prusti2016, Vallenari2023}, that has been used to determine the kinematic parameters of the O--M model. The required input data of the model are positions $\alpha, \delta$, spatial velocities (proper motions $\mu_\alpha\,{\rm cos}\,\delta, \mu_\delta$ and radial velocities V$_r$) of the stars, as well as their distances (parallaxes $\pi$).

\begin{figure}
\centering
\resizebox{\hsize}{!}
   {\includegraphics{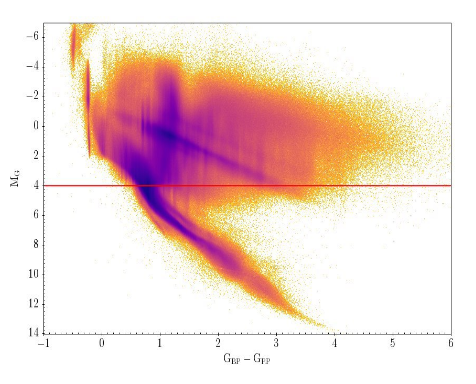}}
  \caption{The color ({\it G}$_{\rm BP}$ - {\it G}$_{\rm RP}$) - absolute magnitude M$_G$ diagram based on the {\it Gaia} DR3 data.}
\label{fig:G-R}
\end{figure}

{\it Gaia} DR3 provides estimates of 5 astrometric parameters (positions, proper motions and parallaxes) for 1.8 billion stars, as well as photometry in {\it G}, {\it G}$_{\rm RP}$ and {\it G}$_{\rm BP}$ bands \citep[][]{Riello2021}. Among them, radial velocities (the sixth astrometric parameter) have been measured for $\sim$33 million objects. The presence of 6 astrometric parameters makes it possible to obtain complete information about stellar motions in space. Of these 33 million objects, we selected  high-luminosity stars with absolute magnitude of M$_G<$ 4 (see Fig. \ref{fig:G-R}) as the base sample, see details in \citet{Akhmetov2023}.

To provide the best quality of astrometric parameters, we excluded from this sample the stars for which the following conditions are satisfied:
\begin{equation}
 \begin{cases}
   RUWE > 1.4,\\
   \pi/\sigma_\pi < 5,\\
 \end{cases}
\end{equation}

About 19.3 million stars meet the criteria listed above and we use them for the kinematic analysis.
Since, as is known from \citet{Lindegren2021}, the {\it Gaia} DR3 parallaxes are systematically biased relative to the expected distribution around zero by several tens of microarcseconds, we corrected them. To correct parallaxes, we used the Parallax bias Z5 computed according to Table 9 given by \citet{Lindegren2021}. Our sample is limited to about magnitude 17 due to the inclusion of radial velocities. Therefore, the magnitude range of the sample has been divided into only two parts. In the range of {\it G} magnitudes from 6 to 13, we applied an offset of -30 $\mu$as, and in the range of {\it G}$>$13, we used a value of -40 $\mu$as. Since in Fig. 20 given by \citet{Lindegren2021} jumps are observed in the range from 11.5 to 13 {\it G} magnitudes, we decided to use an average value of Z5 approximately equal to -30 $\mu$as in this range.

In addition, we corrected for the proper motions of stars in the magnitude range from 6 to 13 as well. \citet{Brandt2018, Lindegren2018} have showed that in the second release of {\it Gaia} (DR2) data, the reference frame of bright stars rotates relative to faint stars and quasars at a velocity of $\sim$0.15 \myr. In the early third {\it Gaia} release EDR3, this rotation has been previously excluded \citep[see section 4.5 by][]{Lindegren2021}. However, an additional correction has been proposed by \citet{Cantat-Gaudin2021}, and we use it to align the stellar proper motions in our sample brighter than {\it G} = 13 with the International Celestial Coordinate System.

\subsection{Distribution of centroids of selected stellar systems in the Galactic plane}

\begin{figure}
\centering
\resizebox{\hsize}{!}
   {\includegraphics{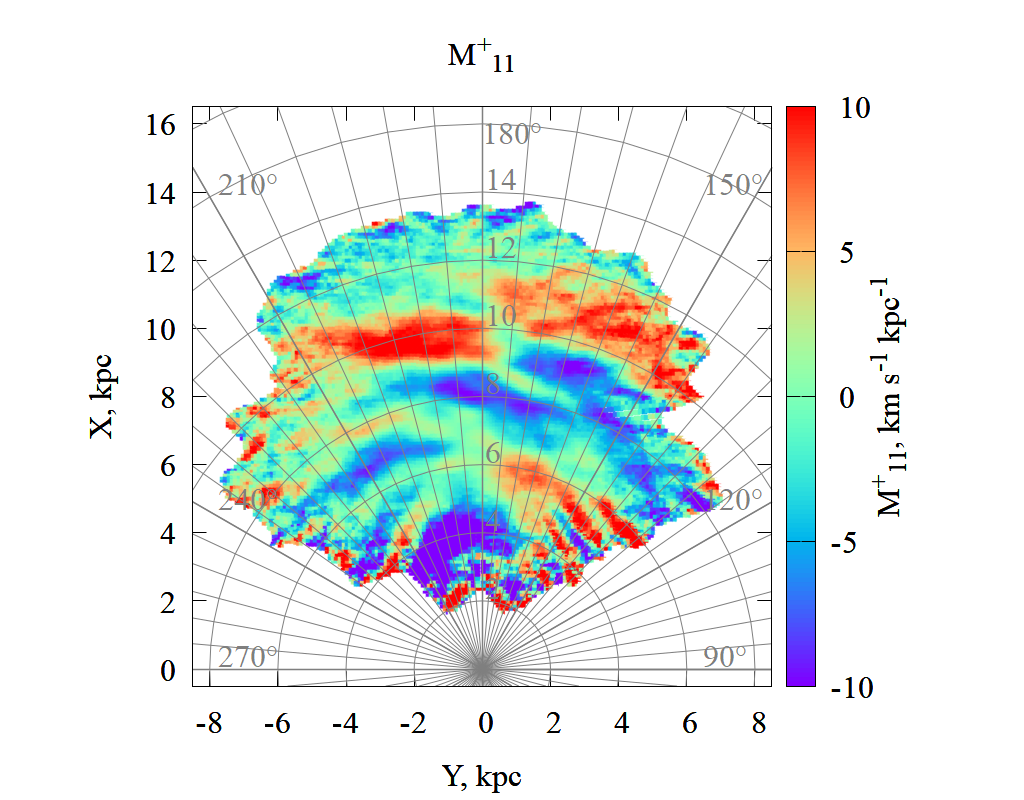}}
  \caption{The O--M model parameter $M^+_{11}$.}
\label{fig:M11_new}
\end{figure}

\begin{figure}
\centering
\resizebox{\hsize}{!}
   {\includegraphics{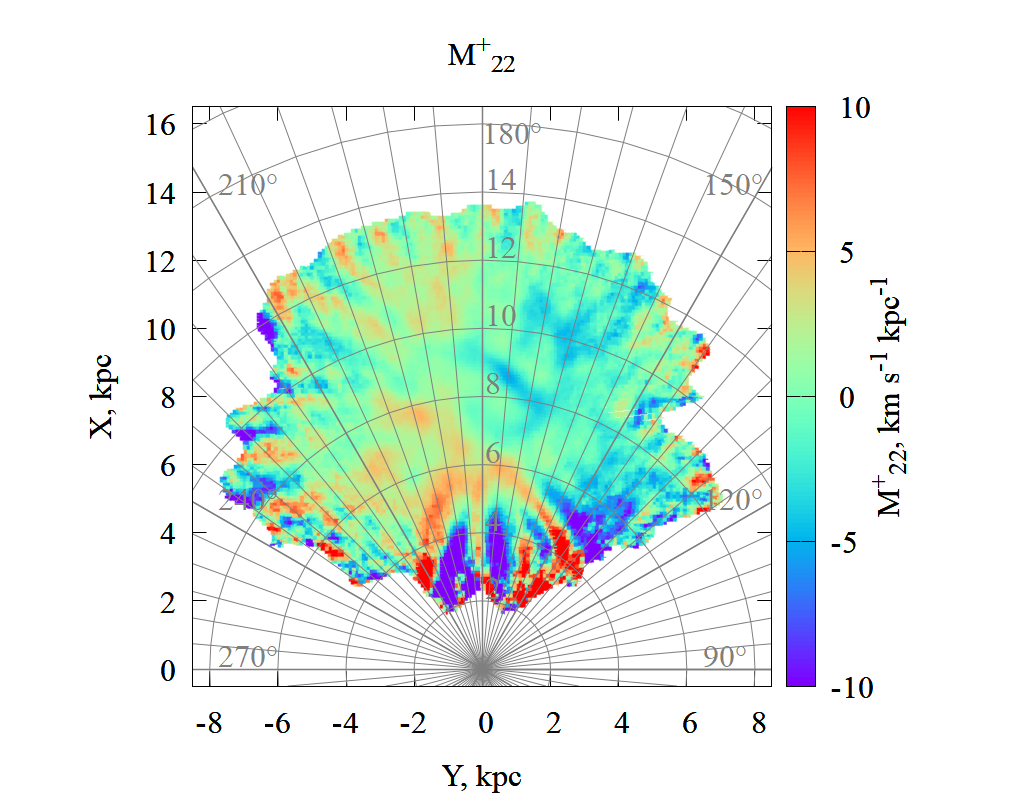}}
  \caption{The O--M model parameter $M^+_{22}$.}
\label{fig:M22_new}
\end{figure}

\begin{figure}
\centering
\resizebox{\hsize}{!}
   {\includegraphics{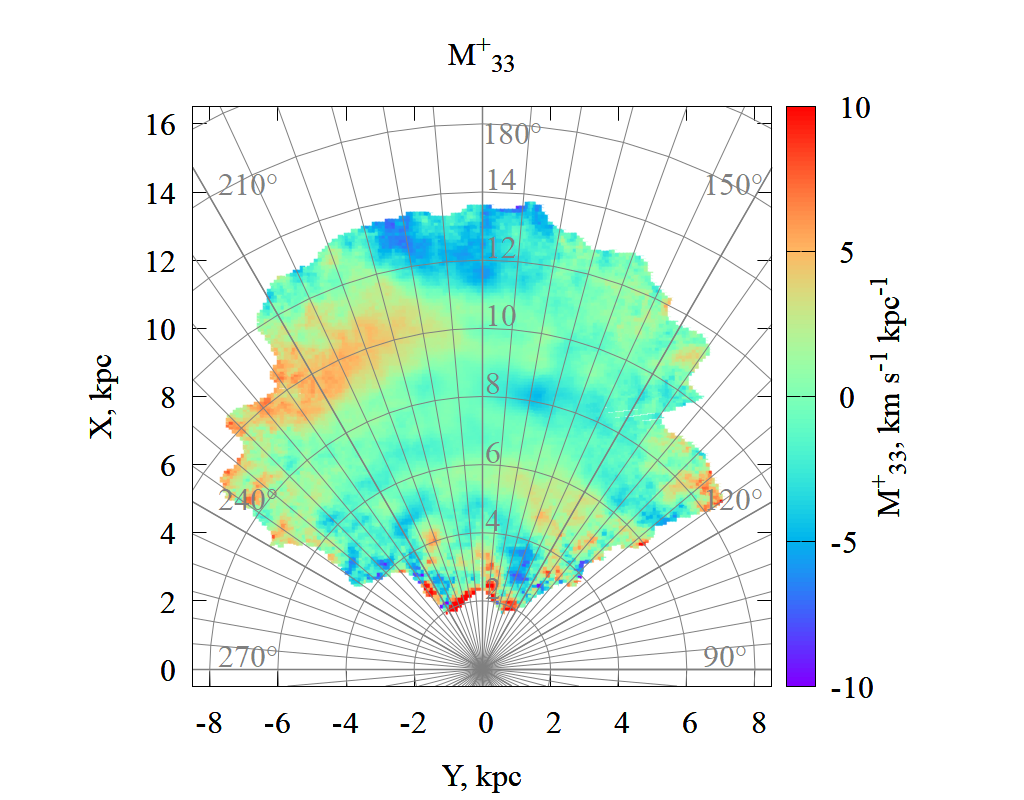}}
  \caption{The O--M model parameter $M^+_{33}$.}
\label{fig:M33_new}
\end{figure}

To solve the system of equations of the O--M model, we divided the final sample into stellar systems, which are regions inside spheres  with a radius of $R_{\rm s}$ = 0.5 kpc. We set the centers of these spheres at the nodes of a rectangular grid given in the Galactic plane. The grid nodes are separated from each other along the $x$ and $y$ coordinates at a distance of 100 pc. We chose the coverage region of the Galactic plane by the rectangular grid of nodes (centroids)  within $140^\circ<\theta<220^\circ$, 1.5 kpc $< R <$ 13.5 kpc of Galactocentric (GC) cylindrical coordinates with distance form Sun to GC is $R_0=8.15 $ kpc(see, for instance, Fig. \ref{fig:M11}, \ref{fig:M22}). We solve the system of equations of the O--M model for 12 unknowns by the least squares method (LSM) using the astrometric data of stars, the number of which in each sphere is at least 1000. Figs. \ref{fig:M11_new}, \ref{fig:M22_new} and \ref{fig:M33_new} show the dependencies of $M^+_{11}, M^+_{22}, M^+_{33}$ on the $X$ and $Y$ coordinates in the range of angles $\theta$ from 210 to 150 degrees and distances $R$ from 4 to 12 kpc.

After calculating the kinematic parameters inside of each sphere, we create a subsample of centroids of those spherical regions in which the value of the parameter $M^+_{11}$ turned out to be insignificant, i.e. the calculated value of the parameter $M^+_{11}$ does not exceed twice the value of its uncertainty. As mentioned above, we do not consider the influence of the parameters $M^+_{22}$ and $M^+_{33}$ on stability, but focus only on establishing a relations between the parameters $M^+_{11}$, which characterize the contraction-expansion of the velocity field in spherical regions and spiral arms. The numerical value of the radius of the spherical region equal to 0.5 kpc has been chosen from empirical considerations and is a compromise between ensuring the reliability of the LSM solution (a sufficient number of stars within the sphere) and detailing of the mapped kinematic parameters.

\begin{figure}
\centering
\resizebox{\hsize}{!}
   {\includegraphics{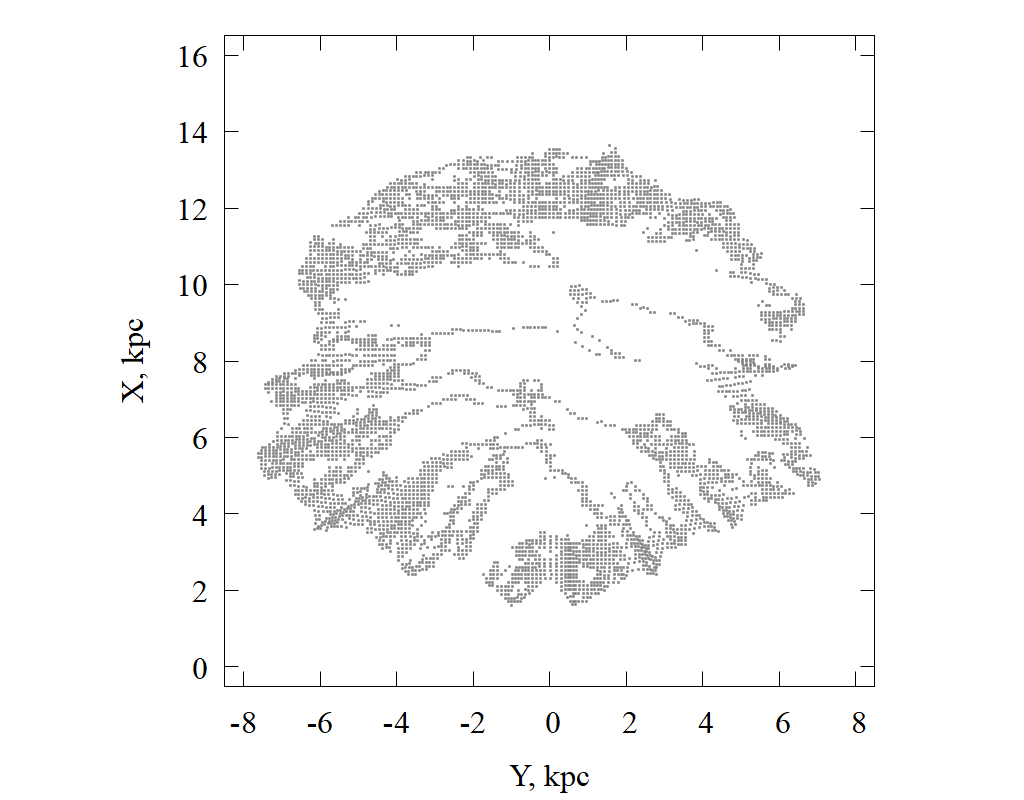}}
  \caption{The distribution of the centroids coordinates in the Galactic $XY$ plane, which are formed from stars contained in spheres with a radius of $R_{\rm s}$ = 0.5 kpc and having insignificant $M^+_{11}$.}
\label{fig:map}
\end{figure}

\begin{figure}
\centering
\resizebox{\hsize}{!}
   {\includegraphics{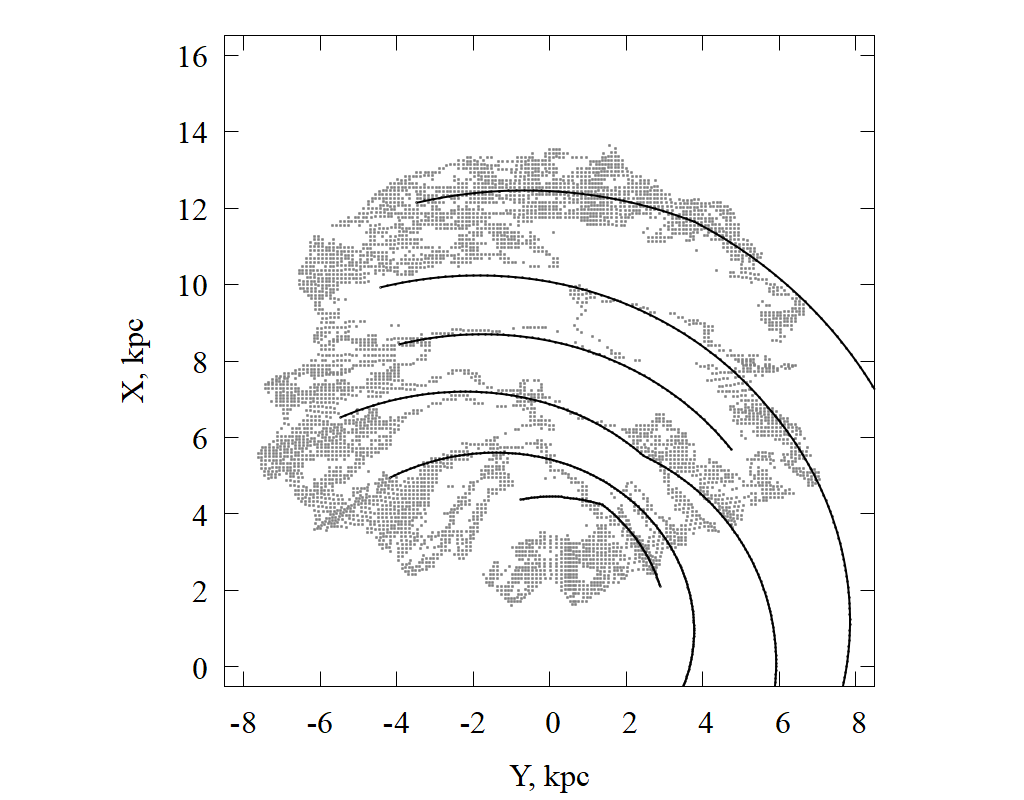}}
  \caption{Same as in Fig. \ref{fig:map} but with superimposed spirals taken from \protect\citet{Reid2019}.}
\label{fig:map_reid}
\end{figure}

Fig. \ref{fig:map} shows the distribution of the centroid coordinates in the Galactic $XY$ plane, which are formed from stars contained in spheres with a radius of $R_{\rm s}$ = 0.5 kpc and having insignificant $M^+_{11}$. To understand whether Fig. \ref{fig:map} contains the centroids belonging to some spiral arms, we superimposed the spiral arms on the centroid distribution with the parameters obtained by \citet{Reid2019}, as shown in Fig. \ref{fig:map_reid}. As can be seen from the latter figure, individual sections of the location of the centroids on the coordinate plane are in good agreement with the superimposed spirals. In this regard, we believe that we can consider the positions of centroids with insignificant $M^+_{11}$ as kinematic tracers of spiral arms in the range of Galactocentric cylindrical coordinates  $140^\circ < \theta < 220^\circ$ , 4 kpc $< R <$ 13 kpc . The choice of these coordinate ranges is mainly due to the accuracy of the parallaxes used and the accuracy of the obtained kinematic parameters. The total number of centroids from the specified range was 25,110. It is these centroids that we use further in the work as kinematic tracers of spiral arms.

\section{Determination and comparison of spiral arm parameters}

To calculate the parameters of the Milky Way spirals, we apply a method that is widely used to study the spiral structure of the Galaxy from the data of various objects \citep[][]{Popova2005, Xu2013, Bobylev2014, Reid2019, Hao2021, Hou2021}. The method is based on the "position angle - logarithm of the distance" diagram.
Considering the centroid as a fictitious star belonging to the spiral arm, the equation that determines the position of this fictitious star on the logarithmic spiral can be written as follows:
\begin{equation}
    R = a_0\,{\rm exp} \{(\varphi - \varphi_0){\rm tg}\,i\}
    \label{eq:log_spiral}
\end{equation}
where $R$ is the Galactocentric distance of the centroid, $a_0 = R( \varphi = \varphi_0 = 0)$ is the Galactocentric distance to the point of intersection of the spiral with the direction of the Galactic center - the Sun, $\varphi$ is the position angle of the centroid, determined from the formula ${\rm tg}\varphi = y/(R_0 - x) $, the value $R_0$ is taken equal to 8.15 kpc, $x, y$ are heliocentric Galactic rectangular coordinates of the centroid, with the $x$ axis directed from the Sun to the Galactic center, and the direction of the $y$ axis coincides with that of rotation of the Galaxy; $\varphi_0$ is some arbitrarily chosen initial angle, which, for example, is taken equal to the average value of all centroid azimuths belonging to the arm. In this paper, we take the angle $\varphi_0$ equal to zero. The pitch angle of the spiral pattern is denoted as $i$ ($i < 0 $ for a leading spiral).

According to our sample, we calculate the heliocentric distance $r$ to a particular centroid from the parallaxes corrected by us, as the average distances to stars included in each spherical region with a radius of 0.5 kpc. We calculate the Galactocentric distance $R$ to a particular centroid and the corresponding position angle $\varphi$ using the following formulas
\begin{align}
    &R^2 = (R_0 - x)^2 + y^2 \\
    &{\rm tg}\,\varphi = y/(R_0 - x)
\label{eq:galactocentric_dist}
\end{align}
where $x, y$ are the heliocentric Galactic rectangular coordinates of the centroid. Obviously, the relations between the position angle $\varphi$ and the coordinate angle $\theta$ is $\varphi + \theta = 180^\circ$.

Since we have chosen $\varphi_0 = 0$, equation \ref{eq:log_spiral} can be rewritten as:
\begin{equation}
    ln(R/a_0) = \varphi\,{\rm tg}\,i
\end{equation}
or in the form:
\begin{equation}
    ln(R) = \varphi\,{\rm tg}\,i + ln(a_0).
    \label{eq:log_r_phi}
\end{equation}
The latter can be rewritten as:
\begin{equation}
    ln(R) = k\varphi\, + b
\label{eq:dist_kb}
\end{equation}
where $k={\rm tg}\,i$, and $b = ln(a_0)$.

As can be seen, relation \ref{eq:dist_kb} is the straight line equation on the plane "position angle - logarithm of distance". The solution of the system of conditional equations separately for each spiral arm by the LSM gives two quantities: $k$ and $b$. Obviously, $i = {\rm arctg} k$, where $i$ is the pitch angle of the spiral, and $a_0 = e^b$ and represents the Galactocentric distance of the point of intersection of the spiral with the $X$ axis directed from the Galactic center and passing through the Sun.

The key point in the approach we use is to find those points (centroids) that belong to a particular spiral arm. Obviously, the choice of such centroids for constructing segments of straight lines on the plane "position angle $\varphi$ - logarithm of the distance ln($R$)" is a rather subjective procedure (see Fig. \ref{fig:map}). We objectified the procedure for selecting points (centroids), which we use later to construct segments of straight lines, by analyzing the behavior of the centroids` radial velocities of our sample depending on the Galactocentric coordinates $R$ and $\theta$.

\begin{figure}
\centering
\resizebox{\hsize}{!}
   {\includegraphics{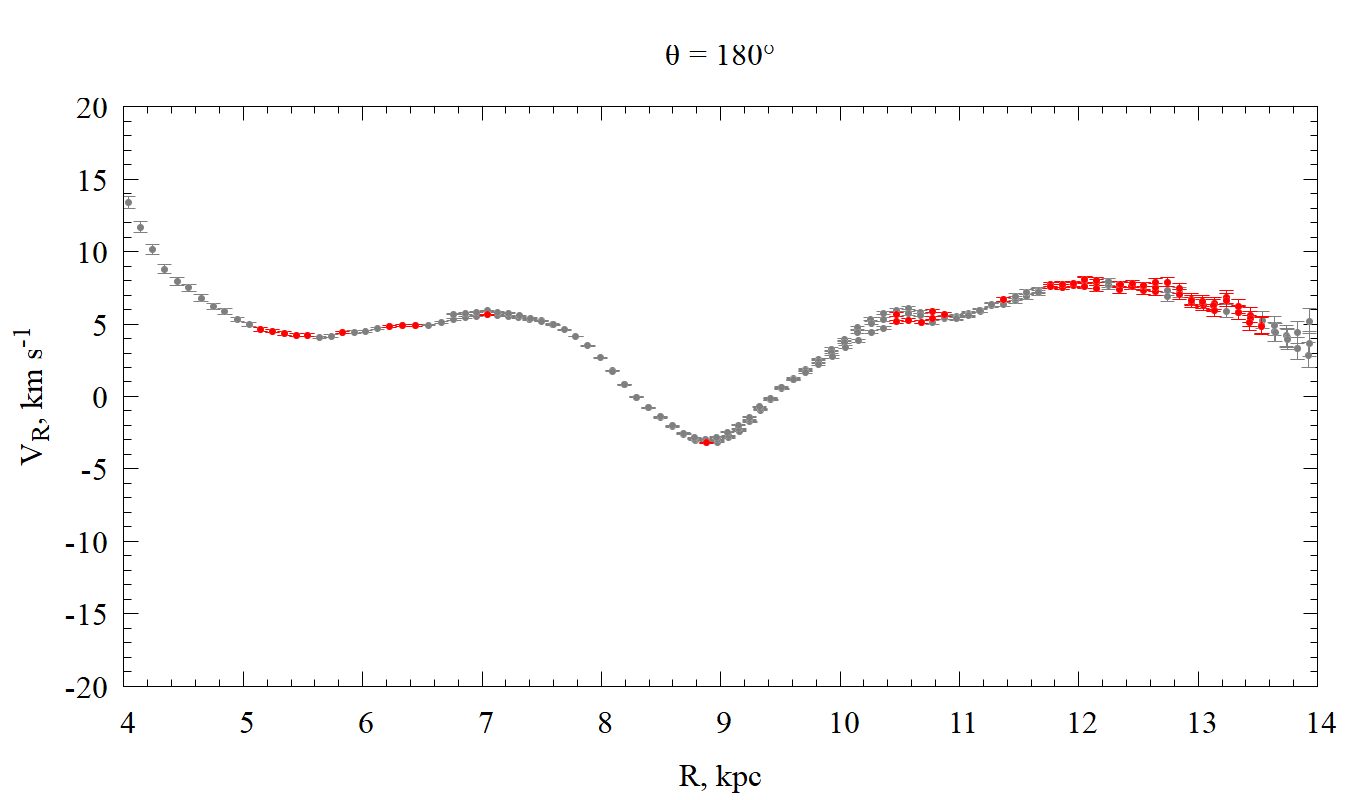}}
  \caption{Dependence of the radial velocities $V_R$ of the centroids of the original sample as a function of the Galactocentric distance $R$ at the fixed angle $\theta = 180^\circ$. Red color shows the centroids for which the condition that the $M^+_{11}$ parameter is not significant has been met.}
\label{fig:Vr_180}
\end{figure}

\begin{figure*}
\centering
\resizebox{\hsize}{!}
   {\includegraphics{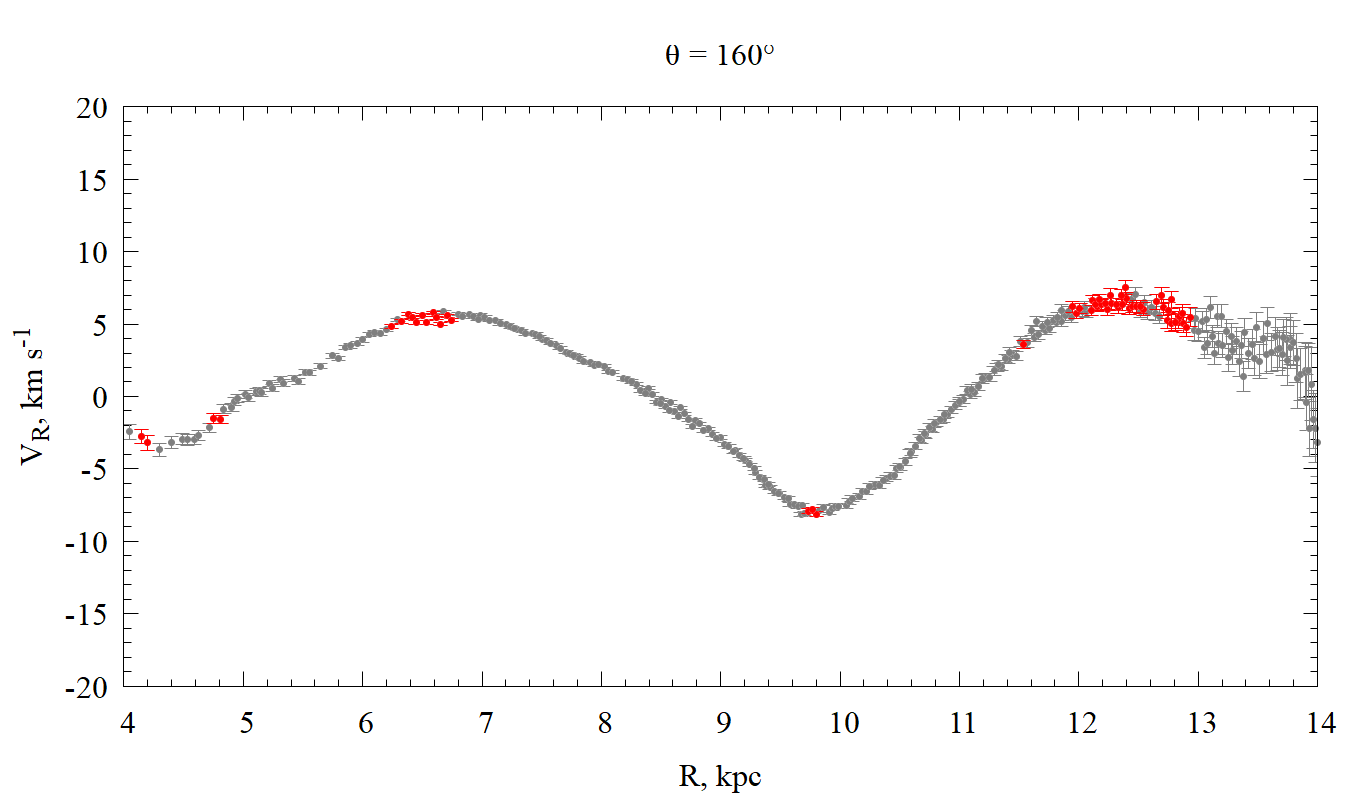}
    \includegraphics{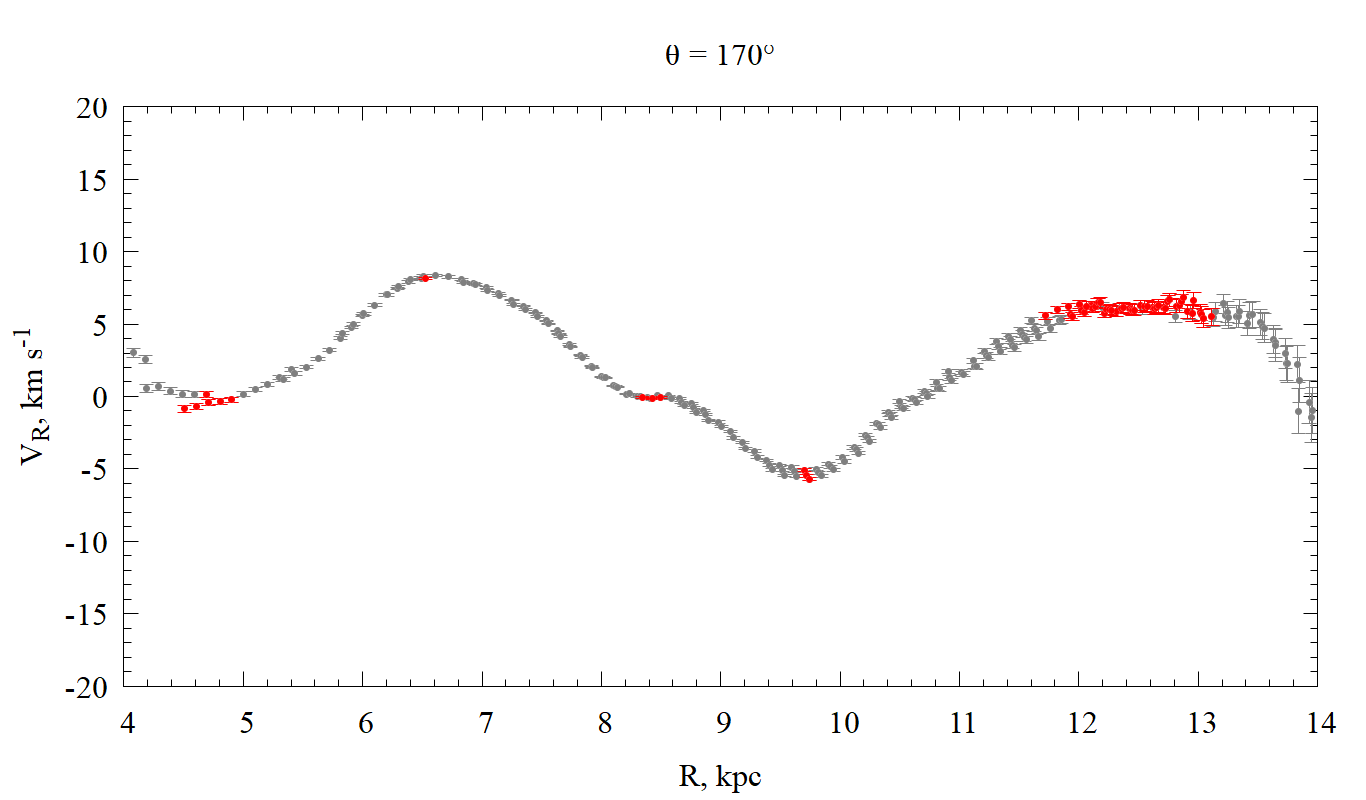}}
\resizebox{\hsize}{!}    
   {\includegraphics{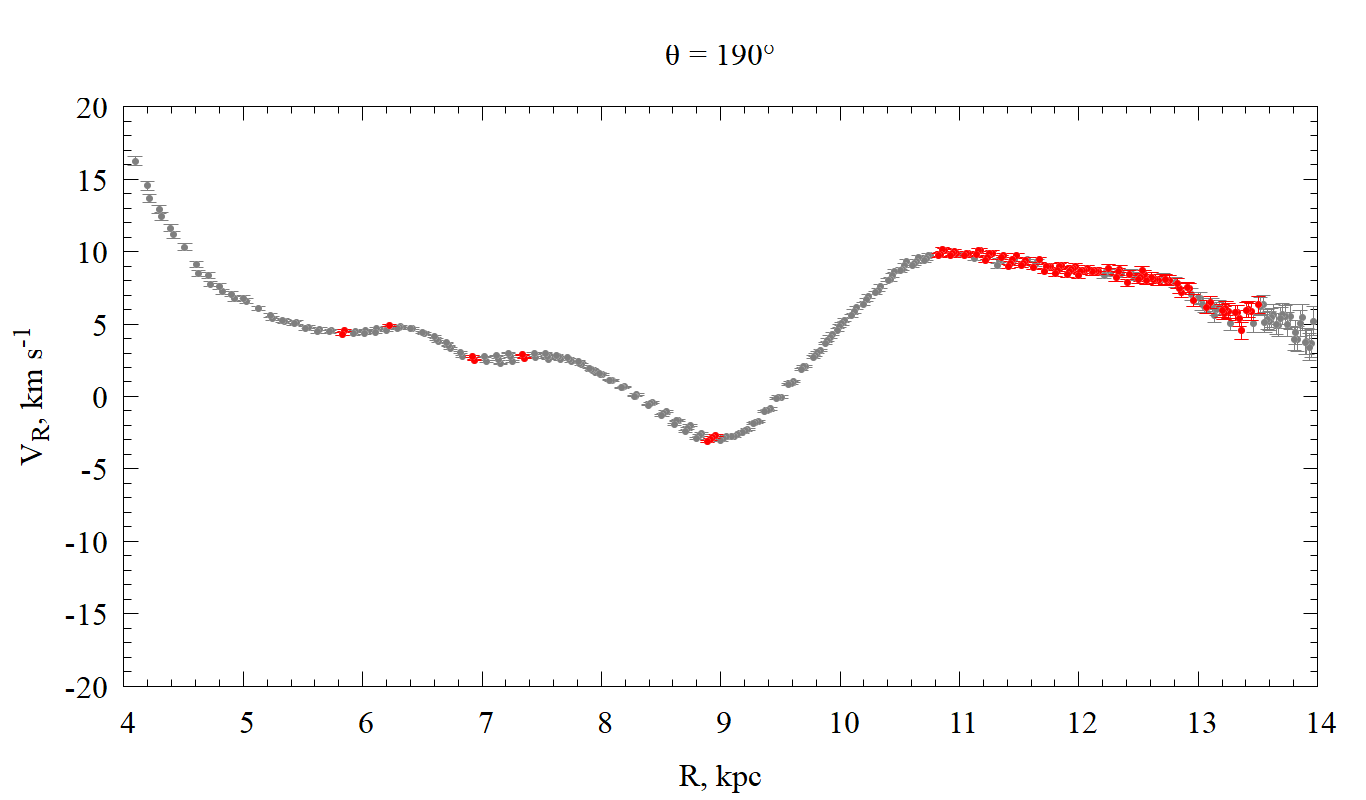}
    \includegraphics{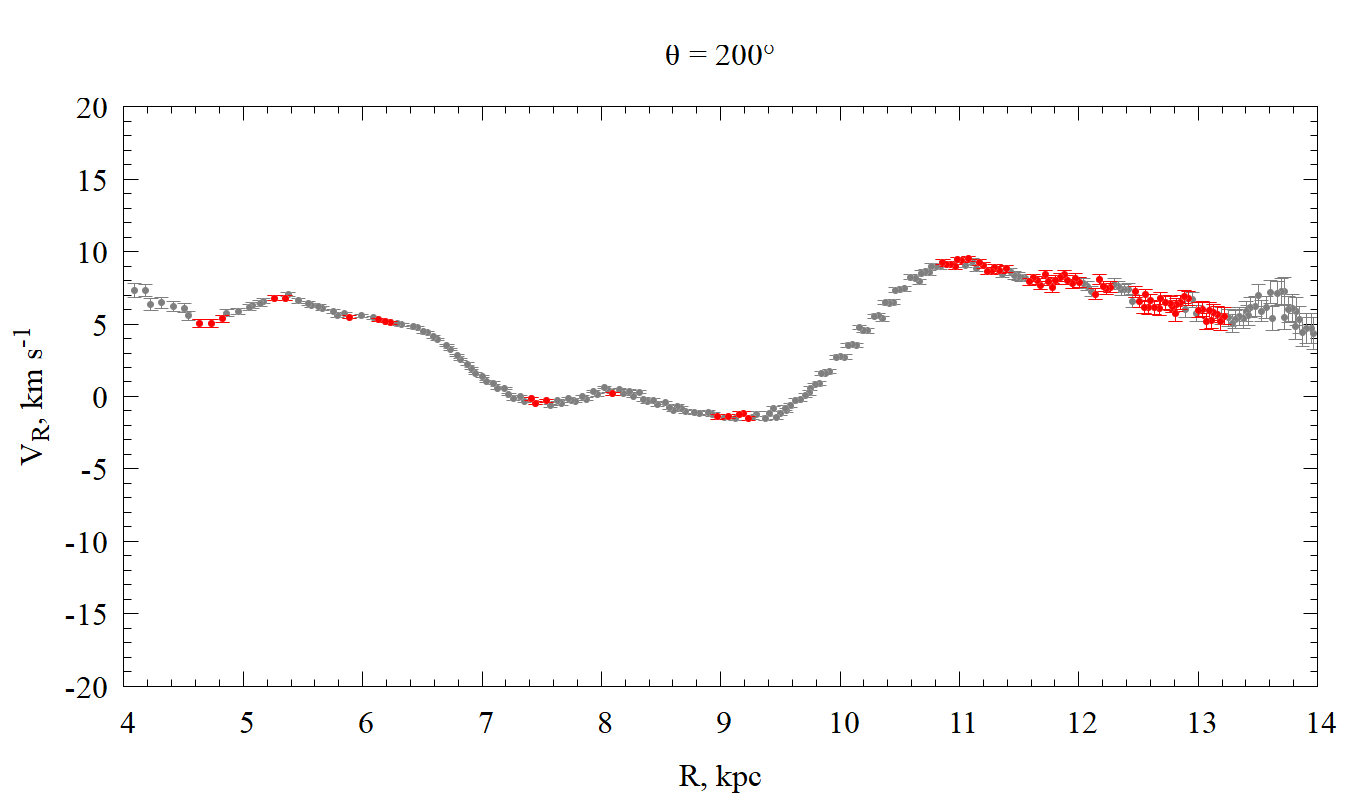}}
  \caption{Same as in Fig. \ref{fig:Vr_180} but at other fixed coordinate angles: $\theta = 160^\circ$ (top left), $170^\circ$ (top right), $190^\circ$ (bottom left) and $200^\circ$ (bottom right).}
\label{fig:Vr_160-200}
\end{figure*}

For example, Fig. \ref{fig:Vr_180} shows the radial velocity of the centroids of the initial sample depending on $R$ at the fixed angle $\theta = 180^\circ$. Fig. \ref{fig:Vr_160-200} shows the same dependencies but at other coordinate angles symmetrically inclined with respect to $\theta = 180^\circ$. We built such dependencies of the radial velocities on $R$ with a step of 1 degree in $\theta$ for the entire part of the Galaxy studied in this work. The red color on these dependencies shows the centroids for which the condition that the $M^+_{11}$ parameter is not significant has been met. As expected, the red dots in the figures are grouped in places of local extrema, as well as in those places where the radial velocity $V_R$ is almost constant.

It is also clearly seen from the figures that when the angle $\theta$ changes, the colored dots move along the $R$ axis in a certain interval $\Delta\,R$. In this case, the average value of the distances $R_m$ from the intervals $\Delta\,R$ turned out to correlate with the Galactocentric distances to the spiral arms $a_0$ known from the literature.
These empirical facts allowed us to form an algorithm for the selection of centroids to construct linear dependencies ${\rm ln}\,R = k\,\varphi + b$.

To form a primary list of centroids belonging to a particular arm, we use the following conditions:

1. In a spherical region with a radius of 0.5 kpc, the $M^+_{11}$ parameter is insignificant: $|M^+_{11}/\sigma_{M^+_{11}}| < 2$.

2. Using the method described above for first approximation values of $a_0$ the mean of Galactocentric distance $R_m$ of the spiral arms turned out to be equal 5.5, 6.8, 8.76, 10.4 and 12.3 kpc for the Scutum-Centaurus arm, the Sagitarius-Carina arm, the Orion local arm, the Perseus arm and the Norma-Outer arm respectively.

3. From the analysis of various works \citep[][and other]{Dambis2015, Reid2019, Veselova2020, Hao2021, Hou2021}, in the first approximation, the pitch angle of all the spiral arms $i$ has been taken equal to -13 degrees.

4. For each spiral arm the values $R_i$ were calculated using the above initial values of $a_0$, $i$ and position angle $\varphi_c$ of the centroids by the formula \ref{eq:log_spiral}.

\begin{figure}
\centering
\resizebox{\hsize}{!}
   {\includegraphics{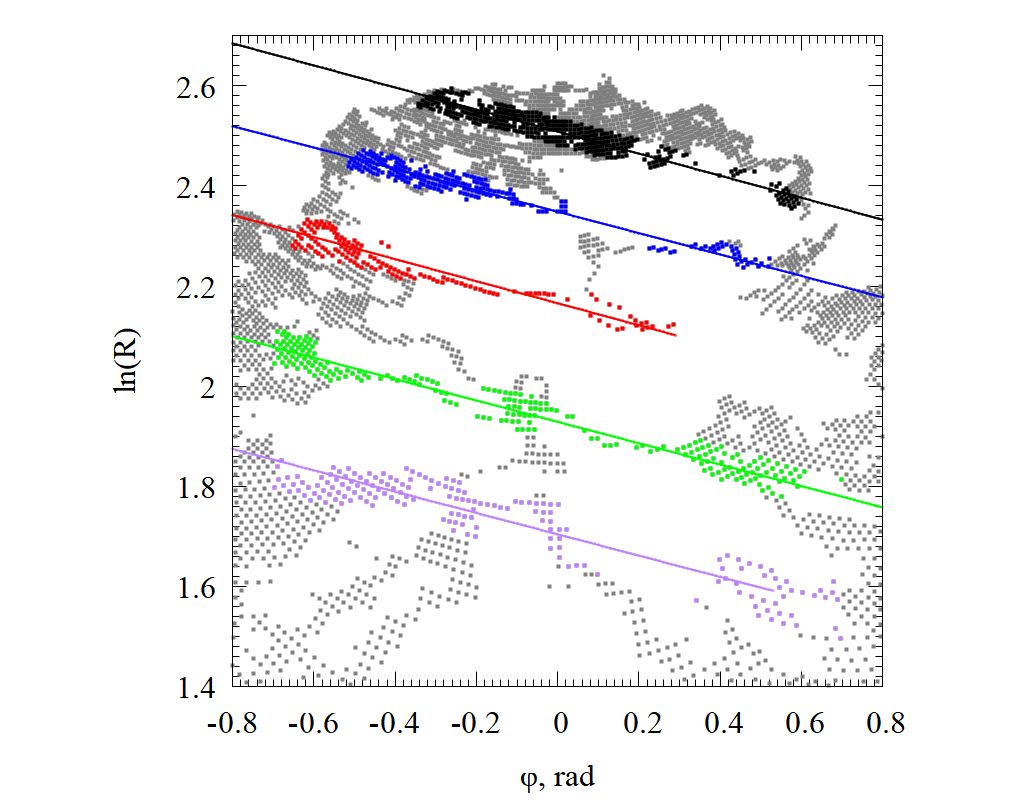}}
  \caption{Dependence of ln($R$) on the position angle $\varphi$. Blue dots and line (\textcolor{blue}{\bf --}) corresponds to the Perseus arm, green dots and line (\textcolor{green}{\bf --}) - to the Sagitarius-Carina arm, black dots and line ({\bf --}) - to the Outer arm, violet dots and line (\textcolor{myviolet}{\bf --}) - to the Scutum-Centaurus arm, red dots and line (\textcolor{red}{\bf --}) - to the Orion local arm.}
\label{fig:lnR_R}
\end{figure}

5. We selected the centroids of the spherical regions with the Galactocentric distance $R_c$ which are in the interval $\pm$300 pc from the value obtained in the previous step $R_i$ for a particular spiral arm, i.e  $|R_i-R_c|<300$ pc as we can see on Fig.\ref{fig:lnR_R}.

We use the $R_c$ and $\varphi_c$ coordinates of centroids of spherical regions satisfying the conditions listed above to solve the equations \ref{eq:dist_kb} by the LSM with iterative excluding centroids according to the criterion: deviation of an individual point from the solution exceeds $2\sigma$.

\begin{figure}
\centering
\resizebox{\hsize}{!}
   {\includegraphics{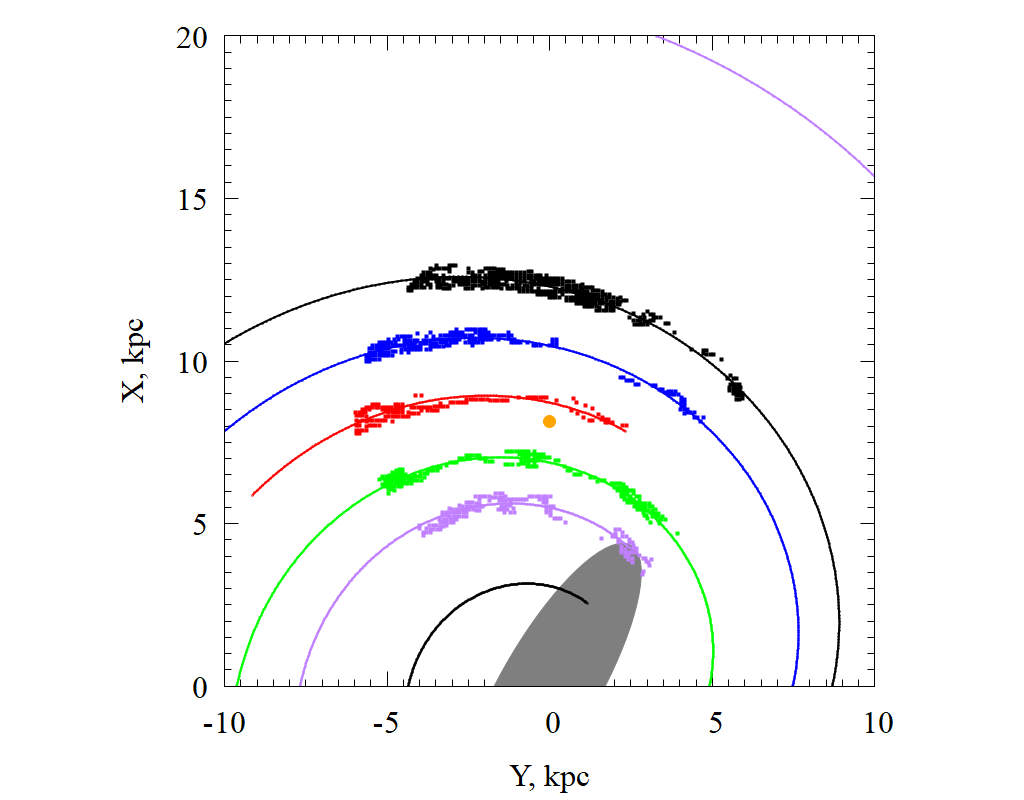}}
  \caption{Logarithmic spirals plotted using the corresponding parameters derived in this work and given in Table \ref{tab:sp_params}. Color points show the centroids belonging to corresponding spiral arms. The gray ellipse shows the Galactic bar reproduced according to the data obtained by \protect\citet{Wegg2015}.}
\label{fig:Arm}
\end{figure}

The final results of determining the parameters of the spiral arms $i={\rm arctg }k$ and $a_0 = e^b$ , as well as their uncertainties, are shown in Table \ref{tab:sp_params} and plotted in Fig. \ref{fig:Arm}.

\begin{table*}
\caption{Parameters of spiral arms. I - Scutum–Centaurus arm, II - Sagitarius-Carina arm, III - Perseus arm, IV - Outer arm, Local - Orion arm. N is the number of centroids used.}
    \centering
    \begin{tabular}{lccccc}
\hline
      & I              & II                & Local             &  III             & IV             \\
\hline
$i$   & -12.04$\pm$0.34   & -12.07$\pm$0.17    & -12.43$\pm$0.32   &  -12.07$\pm$0.18  & -12.43$\pm$0.18   \\
$a_0$ & 5.493$\pm$0.014   & 6.878$\pm$0.010    & 8.719$\pm$0.0.022 &  10.470$\pm$0.012 & 12.289$\pm$0.009  \\
N     & 151               & 202                & 122               &  215              & 371               \\
\hline
\end{tabular}
    \label{tab:sp_params}
\end{table*}
\begin{table*}
\caption{Parameters of spiral arms. (1) - \citet{Bobylev2014}, (2) - \citet{Reid2019}, (3) - \citet{Veselova2020}, (4) - \citet{Hou2021}, (5) - \citet{Xu2018}. I - Scutum–Centaurus arm, II - Sagitarius-Carina arm, III - Perseus arm, IV - Outer arm, Local - Orion arm.}
    \centering
    \begin{tabular}{lcccccc}
\hline
      & I              & II                & Local             &  III             & IV               &Ref               \\
\hline
$i$   & -11.2$\pm$4.0   & -9.3$\pm$2.2    & -10.2$\pm$0.3   &  -14.8$\pm$0.8  & -11.5$\pm$1.9   & 1   \\
$ $ &-13.1$\pm$2.0  &-9.0$\pm$1.9  &-11.4$\pm$1.9  &-9.5$\pm$2.0  &-6.2$\pm$4.2 & 2  \\
$ $ &-11.7$\pm$0.9  &-13.1$\pm$1.4  &-9.9$\pm$1.2  &-6.2$\pm$1.6  &-5.2$\pm$2.8 & 3  \\
$ $ &-8  &-13.8  &-8.5 &-11.9  &     & 4  \\
$ $ &-18.7$\pm$0.8  &-13.5$\pm$0.5  &-11.5$\pm$0.5 &-9.0$\pm$0.1  &     & 5  \\

$ $ \\

$a_0$ & 4.5$\pm$0.2   & 6.8$\pm$0.3    & 8.1$\pm$0.3 &  9.9$\pm$0.4 & 13.5$\pm$0.5  & 1  \\
$ $ & 4.9$\pm$0.1   & 6.0$\pm$0.1    & 8.3$\pm$0.1 &  8.9$\pm$0.1 & 12.2$\pm$0.4  & 2  \\
$ $ & 6.07$\pm$0.04   & 6.78$\pm$0.05    & 8.19$\pm$0.05 &  9.74$\pm$0.09 & 12.02$\pm$0.17  & 3  \\
$ $ &6.5  &6.9  &8.5 &9.6  &     & 4  \\
$ $ &5.9$\pm$0.1  &7.2$\pm$0.1  &8.3$\pm$0.1 &10.6$\pm$0.1  &     & 5  \\
\hline
\end{tabular}
    \label{tab:sp_params_other}
\end{table*}

The derived parameters of spiral arms at the level of accuracy of their determination turned out to be in good agreement with the parameters given in other studies using various tracers (see Table \ref{tab:sp_params_other}).

Below we present plots for linear dependencies on the “position angle - logarithm of distance” plane and plots of logarithmic spirals constructed in the Galactic plane according to the data in Table \ref{tab:sp_params}. Specific dependencies are color-coded. Violet color corresponds to the Scutum-Centaurus arm, green -- to the Sagitarius-Carina arm, red -- to the Orion local arm, blue -- to the Perseus arm, and black -- to the Norma-Outer arm. All figures are built according to the algorithm described above.

\section{Results and discussion}

\begin{figure}
\centering
\resizebox{\hsize}{!}
   {\includegraphics{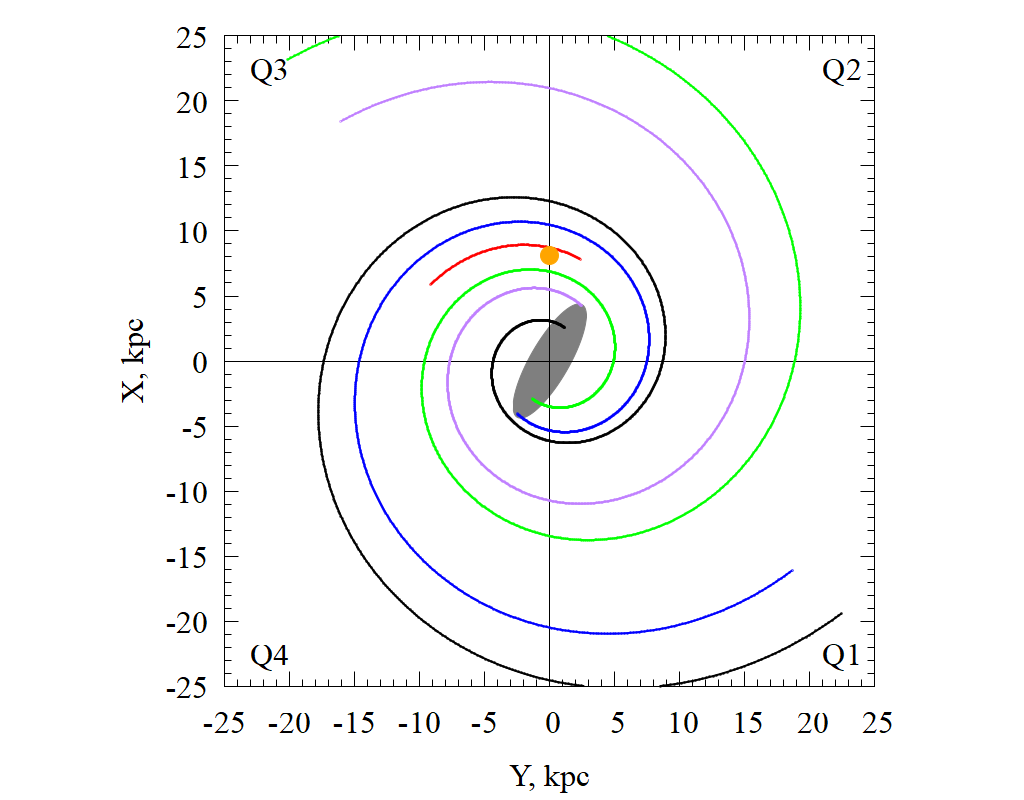}}
  \caption{Schematic representation of the the spiral pattern of the Milky Way in the rectangular Galactocentric coordinate system. Blue curve (\textcolor{blue}{\bf --}) corresponds to the Perseus arm, green curve (\textcolor{green}{\bf --}) - to the Sagitarius-Carina arm, black curve ({\bf --}) - to the Outer arm, violet curve (\textcolor{myviolet}{\bf --}) - to the Scutum-Centaurus arm, red curve (\textcolor{red}{\bf --}) - to the Orion local arm. The yellow point shows the position of the Sun. The gray ellipse shows the Galactic bar reproduced according to the data obtained by \protect\citet{Wegg2015}.}
\label{fig:arms}
\end{figure}

Fig. \ref{fig:arms} shows the four main spiral arms of the Milky Way in the rectangular Galactocentric coordinate system. They are built according to the parameters determined in this work by extrapolation beyond the data we have and schematically represent the spiral pattern of our Galaxy (Grand design). Also, Fig. \ref{fig:arms} shows the Galactic bar reproduced according to the data obtained by \citet{Wegg2015}, where the major and minor axes of the bar were determined to be 5 and 1.5 kpc, respectively \citep[see ][as well]{Rattenbury2007, Cao2013}. Its major axis inclination to the X-axis is approximately 30$^\circ$. The bar is shown in Fig. \ref{fig:arms} schematically as a filled gray ellipse.

Note that, unlike other works, we do not use the concept of "kink" in an arm, introduced by \citet{Reid2019}, but assumed that any arm as a whole can be characterized by one value of the pitch angle of. It is also worth noting that we could not determine the width of the spiral arms in this work. However, it can be recalled that the used tracers of the spiral arms are maximally separated by 600 pc in the Galactocentric distance, which can be considered as a lower estimate of their width.

As can be seen from the figure, the spiral arms intersect with the major axis of the bar at four points. The intersection points of the Norma-Outer and Scutum-Centaurus arms with the major axis of the bar are located in the second Galactocentric quadrant Q2 and have approximate coordinates $x_{\rm N} = 2.5$, $y_{\rm N} = 1.0$; $x_{\rm Sc} = 4.0$, $y_{\rm Sc} = 2.5$ kpc, respectively.
The Norma-Outer arm starts at the indicated point and continues into the quadrants Q3 and Q4, going counterclockwise into the  quadrant Q1. The Scutum-Centaurus arm originates approximately near the point $x_{\rm Sc} = 4.0$, $y_{\rm Sc} = 2.5$ kpc in the quadrant Q2 and extends counterclockwise into the Q3 quadrant and then into the quadrant Q1.

The Sagitarius-Carina and Perseus arms start near the far end of the bar and have approximate coordinates $x_{\rm S} = -3.0$, $y_{\rm S} = -1.5$ kpc; $x_{\rm P} = -4.0$, $y_{\rm P} = -2.5$ kpc, respectively. The Sagitarius-Carina arm from the quadrant Q4 passes counterclockwise through the quadrants Q1, Q2 and Q3. The Perseus arm starts at Q4, passes through Q1, Q2 and Q3, and extends further counterclockwise.

The behavior of spiral arms described above practically coincides with the descriptions of their behavior given in various works to which we refer.

\section{Conclusions}

The parameters of the spiral structure obtained in this work indicate their good agreement with the results obtained by other authors using various indicators of the spiral structure and different approaches for their determination. The use in this work of statistically insignificant kinematic parameters $M^+_{11}$ for the selection of centroids, the positions of which are considered as kinematic tracers of spiral arms, turned out to be a fairly reliable method. 

Note that although the number of centroids used to determine the spiral parameters is a few hundred, the number of stars in spherical regions reaches hundreds of thousands. This allows us to determine the kinematic parameters, as well as the positions of the centroids with high accuracy, which saves us from arbitrariness when choosing tracers for fixing specific spiral arms.

The proposed approach made it possible to find stable structures in the Galaxy, in particular, to reveal spiral arms and determine their parameters. In addition, the method used makes it possible to obtain information not only about spiral arms, but also some additional information. For example, as Figs. \ref{fig:map} and \ref{fig:lnR_R} show, stable structures ($M^+_{11}$ - insignificant) exist not only in the form of spirals, but also in the form of some formations extending in the space between them. The interpretation of these facts requires further accurate study, which is beyond the scope of this article.

\section*{Data availability}
\addcontentsline{toc}{section}{Data availability}
The used catalogue data is available in a standardised format for readers via the CDS (\url{https://cds.u-strasbg.fr}).
The software code used in this paper can be made available upon request by emailing the corresponding author.

\section{Acknowledgements}
This work has made use of data from the European Space Agency (ESA) mission {\it Gaia} (\url{https://www.cosmos.esa.int/gaia}), processed by the {\it Gaia} Data Processing and Analysis Consortium (DPAC, \url{https://www.cosmos.esa.int/web/gaia/dpac/consortium}). Funding for the DPAC has been provided by national institutions, in particular the institutions participating in the {\it Gaia} Multilateral Agreement.

We are immensely grateful to the Armed Forces of Ukraine for the fact that in wartime we still have the opportunity to work and do science.

\bibliographystyle{mnras}
\bibliography{thebib}

\end{document}